\documentstyle{article}
\begin{document}
\title{Star formation in expanding shells}
\author{So\v{n}a Ehlerov\'{a}, Jan Palou\v{s} \\
Astronomical Institute, Academy of Sciences of the Czech Republic,
Bo\v{c}n\'{\i} II 1401, 141 31 Prague 4, Czech Republic}
\maketitle

\begin{abstract}
The star formation induced in dense walls of expanding shells is discussed.
The fragmentation process is studied using the linear perturbation theory.
The influence of the energy input, the ISM distribution and the ISM speed
of sound is examined analytically and numerically. The universal condition
for the gravitational fragmentation of expanding shells is formulated: if
the total surface density of the disk is higher than a certain critical value,
shells are unstable. The critical density depends on the energy of the shell
and the sound speed in the ISM.
\end{abstract}

\section{Introduction}

HI shells (and holes) are structures found in the distribution
of neutral hydrogen of many nearby galaxies.
The density of the gas in their walls is higher than
the average density of the unperturbed ambient medium;
the star formation in rims of HI shells is observed,
e.g. in LMC (Kim et al. 1999), SMC (Stanimirovic et al. 1999),
IC 2574 (Walter \& Brinks 1999), etc.
Its significance for the evolution of galaxies was studied by
Palou\v{s}, Tenorio-Tagle \& Franco (1994) and others.

The gravitational instability of the shell may lead to the formation of 
a cloud and then to the star formation. This instability is
discussed in this contribution. 
The density perturbation on the shell surface is stretched
by the expansion, while its self-gravity supports its growth. The
instantaneous maximum growth rate of the perturbation in the linear 
approximation is (Elmegreen 1994)
\begin{equation}
  \omega = -{3v_{exp} \over R} + \sqrt{{v^2_{exp} \over R^2} +
  \left ({\pi G \Sigma_{sh}
  \over c_{sh}}\right)^2},
  \label{equat1}
\end{equation}
where $R$ is the radius of the shell, $v_{exp}$ is its expansion speed,
$\Sigma_{sh}$ is its column density  and $c_{sh}$ is the speed of sound
within the shell. The  perturbation grows only if $\omega > 0$.

During the evolution, $R$, $v_{exp}$ and $\Sigma_{sh}$ change.
We may ask, which conditions ensure $\omega > 0$  so that the
gravitational instability may develop. One possibility is studied
by Elmegreen (1994).
Shells in galaxies with thin  disks are not spherical but elongated
and even cylindrical.  With the assumption that the diameter
of the cylinder and its height are equal to the height of the galactic
disk, the parameter of the gravitational instability of expanding
shells may be derived:
\begin{equation}
Q_{trigg} = {\kappa c_{ext} \over \pi G \Sigma _{gas}}
\label{elme1}
\end{equation}
where $c_{ext}$ is the speed of sound in the unperturbed
medium and $\Sigma _{gas}$ is the surface density of the
gaseous component of the galactic disk. This relation shows, that the
star formation in expanding rings may be triggered, if the gas surface 
density $\Sigma _{gas}$ is sufficiently high.

In this paper we are interested in studying the fragmentation properties of
HI shells without any assumptions on shapes and dimensions of the
structures. For another application of the fragmentation of expanding
shells see the contribution of G. Parmentier in this book.

\section{Shells in the static homogeneous medium}

During a majority of the evolution the thickness of the blastwave
propagating into the ISM is much smaller than its radius. This is
used by the thin shell approximation, in which the blastwave
is described as the cold infinitesimally thin layer surrounding the hot
medium and expanding. In case of the static homogeneous medium the
blastwave is spherically symmetric and
the analytical self-similar solution can be found (Sedov 1959; Weaver, 
Castor \& McCray 1977). 

The formula (\ref{equat1}) for the
instantaneous maximum growth rate of perturbations shows, that at early stages
of the evolution the shell is stable, as the fast expansion stretches
all perturbations which might appear. Only later, when $v_{exp}$ is small,
$R$ large and $\Sigma_{sh}$
high, the self-gravity begins to play a role and fragments may form.
At the time $t_b$ the growth rate $\omega $ becomes positive.
This is the first moment,
when the shell starts to be unstable and the fragmentation process  begins.

Using the analytical solution we can get
relations for the radius $R(t_b)$, the expansion velocity
$v_{exp}(t_b)$ and $\Sigma_{sh}(t_b)$
at the time $t_b$, see Ehlerov\'{a} et al (1997); or Ehlerov\'a 2000.
The radius $R(t_b)$ is a lower limit to the distance  
on which the fragmentation (and the triggered star formation) may
take place; the expansion velocity $v_{exp}(t_b)$ is an upper
limit to the random component of the velocity of newly created
clouds (or stars).

So far we have assumed the supersonic motion, i.e.
$v_{exp}(t) > c_{ext}$.
If the shell becomes a sound wave before it starts to be unstable
(i.e. before the time $t_b$), it will always remain stable.
It is possible to derive the expression for the minimum energy input, which
is able to trigger the fragmentation process. The condition, that $v_{exp}$ 
at $t_b$ is equal to the sound speed in the ambient medium $c_{ext}$, gives
the critical luminosity $L_{crit}$ (the energy
flux $L$ of the source is constant with time):
\begin{equation}
  L_{crit} = \left ( {c_{ext} \over 8.13\ kms^{-1}} \right )^4
             \left ({c_{sh} \over kms^{-1}} \right )
             10^{51} erg Myr^{-1}
  \label{equat2}
\end{equation}
If the
energy input is greater than $L_{crit}$, the shell starts to fragment;
if the input is smaller, the shell is always stable.
$L_{crit}$ does  not depend on the density of the ambient medium
but it is a strong function of its speed of sound.

\section{Shells in galactic disks}

Numerical simulations must be  used to describe the evolution
of shells in $z$-stratified galactic disks with the gravitational
field.
The thin shell approximation has been applied in numerical simulations
by many authors (see Ikeuchi 1998, for a review on this subject).
The code, which we use in this paper, is an improved and extended
successor to the code of Palou\v s (1990) described in
Ehlerov\'a et al (1997) and in Efremov, Ehlerov\'a \& Palou\v{s} (1999).
The thin shell is divided into a number of elements;
a system of equations of motion,  mass and  energy for each
element is solved and the fragmentation condition (\ref{equat1})
is evaluated.

Numerical simulations include some effects, which are neglected in
the analytical solution. The most important are:
1) the pressure of the ambient medium and radiative cooling;
2) the finite, time-limited energy input from an OB association
(life-time of the source is $\tau $);
and perhaps the most obvious one
3) the stratification of the ISM in galaxies. In this contribution
we  do not take into account
the gravitational field of the galaxy (see Ehlerov\'a 2000, for the
discussion of this subject).

Due to the disk-like ISM density distribution, shapes of expanding shells 
are not spherical, but elongated in the direction of
the density gradient. For higher input energies and thin disks,
large lobes extending to high $z$-distances form. A fraction of
the energy supplied by the hot stars escapes to the galactic halo.
This phenomenon, called the blow-out, decreases the effect of the OB
association on the most dense parts of  shells in the galactic
plane.

Fragmentation properties
vary with the position on the shell. Typically, for shells
growing in a smooth distribution of gas with the $z$-gradient,
a dense ring is created in the region of the maximum density,
which is the most unstable part of the shell, while large lobes
in low-density regions are stable. In the following
we present results for the most unstable part of shells.

The ISM in galaxies is far from being smooth, it is turbulent (see 
a great deal of contributions in these proceedings).
However, coherent shells and bubbles are observed even in a rather
turbulent ISM. Our numerical experiments in a medium with a hierarchy
of density fluctuations show,
that shells are mostly influenced by the large-scale gradients in the
ISM (with the possible exception of the very perturbed regions, where the
coherent shock does not form).
Therefore we think, that the discussion of shell properties
in the smooth medium does not loose its relevance to the real situation.

\section{The critical surface density}

To study the influence of the $z$-stratification,
we simulated shells in different types of the ISM distribution
and different disk thicknesses $n(z)=n_0 \times f({z \over H})$,
where $f$ describes the profile of the disk and $H$ is its thickness.
We do not assume any relation between $H$, $n_0$ and $c_{ext}$;
we fix the velocity dispersion in the shell $c_{sh}$ to a constant value,
$c_{sh} = 1\  km s^{-1}$.
Results of calculations are as follows:
\begin{enumerate}
 \item{For a given total energy $E_{tot}$ and the sound speed $c_{ext}$
       shells are unstable, if the surface density of the gaseous disk exceeds
       a certain value. 
      }
 \item{The minimum gas surface density of the disk needed for the onset of
       the fragmentation does not depend on the $z$-profile of the disk.
      }
 \item{The value of the this critical surface density depends on the
       $E_{tot}$ and $c_{ext}$.
      }
 \item{Exceptions to these three rules appear for i) very small shells,
       which are to a high degree spherical, and therefore does not
       feel the influence of the disk $z$-stratification; and ii) shells with
       a very profound blow-out, which enables an escape of a substantial
       part of the energy to the halo.
      }
\end{enumerate}

From simulations we derive the fit for the critical surface density
of the gaseous disk:
\begin{equation}
\Sigma_{crit}  =  0.27 \left ({ E_{tot} \over 10^{51} erg}
\right )^{-1.1} \left ({c_{ext} \over km s^{-1}}\right )^{4.1}
10^{20} cm^{-2}.
\label{fit}
\end{equation}

\section{Conclusions}

There exist a critical value of the disk gas surface density 
$\Sigma _{crit}$,
shells expanding in disks with the lower gas surface density are stable,
shells expanding in disks with the higher density are
unstable, can fragment and form gaseous clouds and stars.
The value of $\Sigma_{crit}$ depends on the $E_{tot}$ and $c_{ext}$.
How does it agree with other criteria for spontaneous or triggered
star formation?

The Q criterion for the radial instability of the gaseous disk
\begin{equation}
Q_{sp} = {\kappa c_{ext} \over \pi G \Sigma }
\label{equat3}
\end{equation}
(Safronov 1960), or similar criterion for stellar particles by Toomre
(1964), predicts, that the critical surface density is linearly
dependent on the sound speed. Similarly, the Elmegreen criterion
(\ref{elme1}) for the fragmentation of expanding shells shows
the same behavior. The critical surface density (eq. \ref{fit})
derived by us depends on a fourth power of the sound speed in the ISM.
The main difference is, that we do not study effects of the gravitational
field of the galaxy (which leads to the dependence on the epicyclic
frequency $\kappa$), but instead we ask, that the expansion of
the shell must be supersonic,  otherwise the evolution of instabilities
is not described well by the equation (\ref{equat1}), and moreover 
(and more importantly)
the accumulation of mass by the blastwave is stopped and the growth
of density perturbations is significantly slower or inhibited.

Values of $\Sigma_{crit}$ for reasonable values of $E_{tot}$ and
$c_{ext}$ are
of the order of $(10^{20} - 10^{21})\ cm^{-2}$, which coincides
with the value of observed threshold surface densities for the
star formation in galaxies (Kennicutt 1997; Hunter, Elmegreen
\& Baker 1998).
The agreement of these two quantities may indicate, that the
contribution of the star formation induced by shells to the total
star formation may be important.

The very steep dependence of  $\Sigma_{crit}$ on $c_{ext}$, as
given by the equation (\ref{fit}), indicates  the importance of the
self-regulating feedback for the triggered star formation mode.
Young stars in OB
associations release the energy and compress the ambient ISM, creating
shells. In dense walls of shells, the star formation may be triggered,
if the disk surface density surpasses a critical value
$\Sigma_{crit}$.
The star formation is accompanied by the heating of the ISM, i.e.
increasing $c_{ext}$. This leads to
the increase of $\Sigma_{crit}$ and a subsequent
reduction of the star formation rate. The energy dissipation and cooling
of the ISM decreases  $c_{ext}$ and $\Sigma_{crit}$ and
increases  the star formation, closing the self--regulating cycle
of the triggered mode of the star formation.

The Q criterion (\ref{equat3}) describes the spontaneous
mode of the star formation. Our condition (\ref{fit}) describes the
triggered star formation (and even more precisely, the self-propagating
star formation). Then
the less steep dependence of
$\Sigma_{crit}$ on $c_{ext}$ for the spontaneous star formation
means, that this mode may keep its effectivity in regions, where $c_{ext}$ 
has been increased and the triggered mode of star formation
has been stopped.

{\small
Authors gratefully acknowledge a financial support by the Grant
Agency of the Academy of Sciences of the Czech Republic under the grant
No. A300305/1997 and
a support by the grant project of the Academy of Sciences of the
Czech Republic No. K1-003-601/4.
}


\begin{thebibliography}{}
\bibitem{} Efremov Yu.N., Ehlerov\'a S., \& Palou\v s J. 1999, AA , 350, 457
\bibitem{} Ehlerov\' a S. 2000 PhD Thesis, Charles University, Prague
\bibitem{} Ehlerov\' a S., Palou\v s J., Theis Ch., \& Hensler G.
1997, AA , 328, 121
\bibitem{} Elmegreen B. G. 1994, ApJ , 427, 384
\bibitem{} Hunter D.A., Elmegreen B. G., \& Baker A.L. 1998, ApJ , 493, 595
\bibitem{} Ikeuchi S. 1998, in The Local Bubble and Beyond (IAU Colloquium
No. 166), eds. D. Breitschwerdt, M.J. Freyberg, J. Tr\"umper,
Springer-Verlagp. 399-408
\bibitem{} Kennicutt, R.C. 1997, in The Interstellar Medium in Galaxies, ed.
J. M. van der Hulst, Kluwer Academic Publishers, p. 171
\bibitem{} Kim ~S., Dopita ~M.~A., Staveley-Smith ~L., \& Bessell ~M.~S.
1999, AJ , 118, 2797
\bibitem{} Palou\v s J. 1990, in The Interstellar Disk-Halo
Connection in Galaxies, ed. H. Bloemen, Sterrewacht Leiden, The Netherlands,
p. 101
\bibitem{} Palou\v{s} J., Tenorio-Tagle G., \& Franco J. 1994, MNRAS , 270, 75
\bibitem{} Safronov V. S. 1960, Annales d\' Astrophysique 23, 979
\bibitem{} Sedov L. 1959, Similarity and Dimensional Methods in Mechanics,
Academy Press, New York
\bibitem{} Stanimirovic ~S., Staveley-Smith ~L., Dickey ~J.~M., Sault ~R.~J.,
\& Snowden ~S.~L. 1999, MNRAS , 302, 417
\bibitem{} Toomre A. 1964, ApJ , 139, 1217
\bibitem{} Walter F., \& Brinks E. 1999, AJ , 118, 273
\bibitem{} Weaver R., McCray R., \& Castor J. 1977, ApJ , 218, 377
\end{thebibliography}
\end{document}